\documentclass[11pt]{article}
\usepackage{graphicx}

        \title{Optical element for  X-ray microscopy}
        \author{G.Chadzitaskos\\
Phys. Dept., Faculty of Nuclear Sciences and Physical Engineering,\\
Czech Technical University in Prague, B\v rehov\'a 7, CZ - 115 19 Prague,\\
e-mail: goce.chadzitaskos@fjfi.cvut.cz}
        \begin{document}
        \maketitle

 \begin{abstract}
We present a proposal for a X-ray optical element suitable for X-ray
microscopy and other X-ray-based display systems.  Its principle is
based on the Fresnel lenses condition and the Bragg condition for
X--ray scattering on a slice of monocrystal. These conditions are
fulfilled simultaneously due to a properly machined shape of the
monocrystal with a stress at its ends.

 \end{abstract}


\section{Introduction}

X-ray radiation is electromagnetic radiation with wavelengths
shorter than 10 nm. X--ray microscopes have not used  optical lenses
which operate in visible part of light spectrum. Instead X-ray
microscopy uses as display elements Fresnel zone plates, biconcave
refractive X lenses, polycapillar optics or G\"obel mirrors
\cite{P-W-G}.

Fresnel zone plates are Fresnel lenses produced by means of electron
lithography. Lithographic methods (but other manufacturing
procedures are possible as well) can be used to create ringlets,
 or parts of concentric paraboloid rings, that ensure that radiation emitted
 from a point on the ringlets'
central axis (optical axis), having passed through the ringlets, are
focused back again into a point lying on this axis. The thickness of
these ringlets, or their parts, is in orders of tens of nm and their
height is roughly 1000 nm. These zonal plates are used as radiation
condensers for scanning and display microscopes; in display
microscopes they are used also as lenses \cite{M-F,B-D}.

Another possible way to produce Fresnel zone plates is to create a
set of little plates with separate segments containing parallel
strips with identical spacing between each other. The diameters of
such optical elements are not greater than several centimeters
\cite{Ch-H,S-A-H}.

Biconcave refractive X lenses take the advantage of the fact that
the refractive index of certain materials for X-ray radiation is
only slightly higher than one. A set of identical cavities, located
one after another, creates a condensing lens with long focal
distance. In practice, two crossed systems of biconcave lenses are
often used, each focusing the radiation in one direction only. The
result is a focusing into a point, which, owing to little
differences between the refractive index of the matter and the air,
is located in relatively long distance from the lenses - in order of
meters \cite{Len}.

G\"obel mirrors are plates formed by alternated thin layers of two
metals, such as wolfram and silicon, with a shape of part of the
surface of elliptic spheroid. These mirrors focus monochromatic
X-ray radiation into a point.

Polycapillar optics is created by a set of curved capillaries, which
lead the X-ray radiation and focus it into a point.

All optical elements described above feature big ratio of focal
distance to their diameter. This result in relatively long focal
distances and small diameters the production of which is costly.
They are used in X-ray microscopy for normal as well as so-called
scanning microscopes. In scanning microscopes, the beam is focused
by the optics into a point on a specimen and a detector senses
intensity of X-ray radiation, which has passed through. By measuring
all points - which is carried out by shifting the specimen in two
mutually orthogonal directions - the specimen's final image in given
wavelength is obtained. Information about intensity of the radiation
from the X-ray radiation detectors or digital sensors, such as CCD
cameras, which may be used instead of detectors, is stored in a
computer where the final image is processed.

All examples described above result in relatively high absorption
rate of X-ray radiation and all those systems also feature
relatively long focal distances, i.e. distances between the display
and object, and they can be used for longer wavelengths, around 10
nm. In addition, these components impose high demands on production
equipment and therefore they are rather expensive.

\section{Optical Element}

Some properties mentioned above are improved by an optical set-up
exploiting a monocrystal, which displays X-ray radiation with
wavelength $\lambda$ according to the presented solution. This
set-up consists of at least one monocrystal with atomic planes in
parallel with the optical axis, which is a line connecting the point
to be displayed with the center of its image.The mutual distance of
planes in resting state is $d_0$. The cross section of the
monocrystal is variable. With respect to the optical axis of the
farther or closer side of monocrystal, orthogonal to this optical
axis, is equipped with a device to create a pull or push force and
to maintain a pull or push strength in direction orthogonal to
atomic planes of such monocrystal. The size of cross section S, of
the monocrystal at a distance R from optical axis, is directly
proportional to pre-selected cross section $S_0$ and the push or
pull force F.The cross section is also indirectly proportional to
modulus of elasticity E of given monocrystal in the direction of
force in action. The monocrystal's cross section is given by the
following formula

$$ S=\pm \frac{F}{E}(\frac{1}{\frac{n \lambda}{2 R d_{0}}
\sqrt{R^{2}+s^{2}}-1} ),$$

and the force F is defined by the equation

$$ F=\pm S_{0}E( \frac{n \lambda}{2 R_{0} d_{0}} \sqrt{R_{0}^{2}+s
^{2}}-1),   $$

where s, with respect to the monocrystal's longitudinal axis, is an
object and also a display distance and $S_0$ is a pre-selected cross
section of monocrystal upon requirements of the application in
distance $R_0$ from the optical axis and n is a natural number. In
both equations, the + sign applies for pull force, the - sign
applies for push force. Further, as already mentioned above,
$\lambda$ is the X-ray radiation wavelength and $S_0$ is a
pre-selected cross section in distance $R_0$.

In another possible embodiment, the optical set-up is formed by
minimum two identical monocrystals arranged around the optical axis.
All such monocrystals feature equal object and display distance s,
while each monocrystal is equipped with a device applying push or
pull force F within the minimum distance $R_0$ from the optical axis
and the maximum distance $R_m$ from the optical axis. Their cross
section is calculated by means of the above-mentioned equation. In
other words, it is possible to arrange the monocrystals around the
optical axis provided their object and display distances for the
same wavelength are equal.

 An advantage of this proposal is the relatively simple production of its
components, which is based on Hooke's law without needing to know
the exact position of individual atomic planes. Production of the
current Fresnel structures involves electron lithography for
individual ringlets. This method, instead requires only precise
machining of the outer shape, for instance cutting using water jet
or a laser, and applying a force, which can be controlled. Machining
of the outer shape results in better physical effect, it means lower
absorption and larger active area, with lower costs because
lithography is significantly more expensive than cutting. Another
advantage is that it works for also for lower order wavelengths,
i.e. below 1 nm.

\section{Examples}

An example of the optical set  set-up designed by the presented
proposal is shown in attached drawings. Fig. 1 shows examples of
creating two different monocrystals and of basic displaying set-up.
Fig. 2 shows utilization of the set-up for rotational X-ray scanning
microscope. Fig. 3 shows a method of creating an optical element
from multiple monocrystals and fig. 4 shows a chart of dependence of
sections ratio $S/S_0$ on distance from optical axis used for
determination of pull force on atomic planes.

Fig. 1 a) and b) show two examples how monocrystal 1 can be created,
with designated atomic planes 2. a) schematically shows possible
shape of monocrystal 1, an asymmetric shape, designed for shifting
the atomic planes by push force, while Monocrystal's 1 shape b) is
designed for shifting the atomic planes by pull force. In both
cases, the atomic planes 2 where the beam reflects or refracts are
horizontal. The first example shows a monocrystal 1 where optical
axis, i.e. a line connecting a point to be displayed with center of
its image, is located under the monocrystal's 1 bigger cross section
and the arrow indicates a push force applied by push device. On the
contrary, b) shows situation when the monocrystal's 1 imaginary
optical axis is located under the monocrystal's 1 smaller cross
section and the arrow indicates a pull force applied by pull device.
Push and pull devices may be implemented in various ways, such as a
fixed point and piezoelectric crystal, screw mechanism and tension
gauge, etc., and they can be located on one or both opposite sides
of monocrystal 1. c) shows example of an optical set-up displaying
X-ray radiation with wavelength $\lambda$. This set-up consists of
monocrystal 1 with atomic planes 2, which are arranged in parallel
with optical axis 3. Mutual distance of atomic planes 2 in resting
state is $d_0$ and the monocrystal's 1 cross section S is variable.
On the farther side of monocrystal 1 from the optical axis 3,
orthogonal to this optical axis 3, the cross section is smaller than
on the side closer to the optical axis 3, and therefore it is
equipped with a push device, which is not shown in the figure.
Purpose of this push device is to maintain a push force F in
direction orthogonal to atomic planes 2 of this monocrystal 1.

The figure schematically shows function of symmetrically processed
monocrystal 1, exposed to push force, which focuses X-ray radiation
with wavelength $\lambda$ from distance s to distance s. Both these
distances must be equal. Monocrystal 1 is located between distances
$R_0$ and $R_m$ from optical axis 3 where $R_0$ is the minimum
distance and $R_m$ is the maximum distance from optical axis 3.
Based on given equation, cross section S is a function of distance
R. Subsequently, the dependence of cross section S on distance R is
additionally calculated according to the given equation based on
calculated force F and modulus of elasticity E. A plate with shape
complying with the equation is cut from monocrystal 1. Rays emitted
from a point at the object impact the monocrystal 1, reflect from
individual atomic planes 2 and are projected in a display plane as
shown in the figure.

In case of rotating X-ray scanning microscope with a source 4 of
X-ray radiation, single monocrystal 1 may be used, embodied as shown
in fig. 2, where the specimen 5 rotates around axis orthogonal to
it, fixed in space, passing through the center of specimen 5. Such
specimen 5 is after each revolution gradually shifted in direction
of the arrow, thus allowing its scanning from one end to the other.
Intensities of radiation measured by radiation detector 6 are
processed by computer into the resulting image similarly as in case
of tomography. Simultaneously, this monocrystal 1 acts a
monochromator and a separate monochromator may thus be excluded.

Optical set-up may consist of multiple identical monocrystals 1,
which are arranged around the optical axis 3 as shown in fig. 3. All
these monocrystals 1 have equal object and display distance, while
each monocrystal 1 is equipped with a device applying on each
monocrystal 1 pull force F between the minimum distance $R_0$ of the
bottom side of monocrystals 1 from optical axis 3 and between
maximum distance $R_m$ from optical axis 3. Their cross section is
again calculated according to the given equation. Shape of
monocrystals 1 may vary provided their pre-selected cross sections
are maintained. In this case, as an example, an optical element
consisting of 28 monocrystals 1 is shown, where atomic planes 2 act
directly as a set-up of concentric polygon ringlets. By selecting
suitable shape and applying suitable force in compliance with the
equation it is possible to achieve a situation when among the atomic
planes 2 are such distances, which cause the same constructive
interference as Fresnel lens for selected wavelength, or possibly
also for wavelength, which is half of such wavelength, its third,
etc., and act simultaneously as a monochromator, which selects only
these wavelengths from the whole spectrum used. The same effect may
be achieved by using suitable gradient of admixtures in crystal,
such as hydrogen in gadolinium, which affects interatomic distance.
Implementation of detectors is the same as in other cases.

Fig. 4 shows chart with dependencies of sections ratios $S/S_0$ of
monocrystal 1 on distance from optical axis 3 used for determination
of pull force on atomic planes 2 for the following values: $\lambda$
= 0.1 nm, $ d_{0}$ = 0.4 nm, $R_{0}$ = 0.01 m, s = 0.5 m and n = 1,
in the extent up to $R_{m}$ = 0,07 m.

\section*{Acknowledgements}
The authors wish to acknowledge support of the Ministry of Education
of Czech Republic under the research project MSM6840770039.

\newpage

\begin{figure}[h]
\begin{center}
\includegraphics[width=3in]{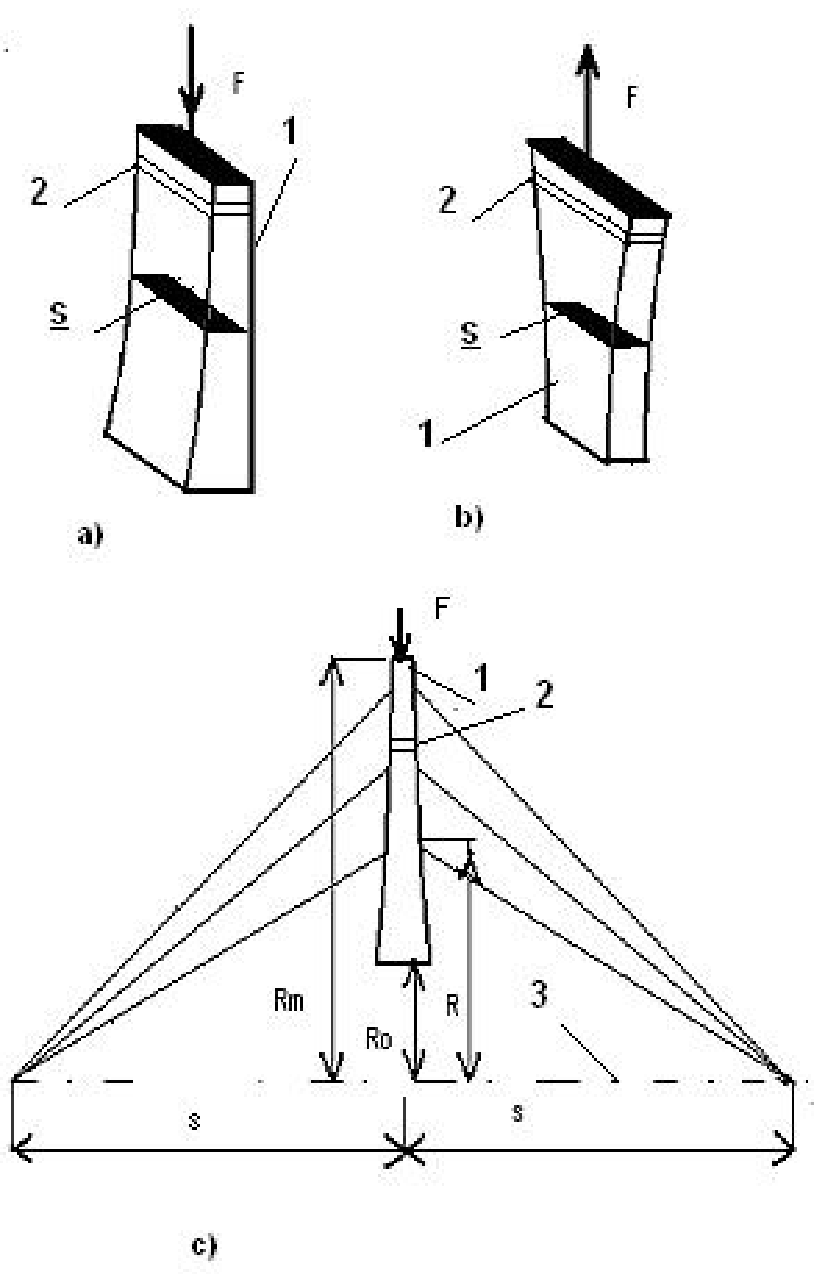}
\end{center}
\caption{ a) and b) show two examples of  possible shape of
monocrystal 1, designed for shifting the atomic planes 2 by push and
pull force F. c) shows example of an optical set-up displaying X-ray
radiation with wavelength $\lambda$. It consists of monocrystal 1
with atomic planes 2, which are  parallel with optical axis 3.
Mutual distance of atomic planes 2 in resting state is $d_0$ and the
monocrystal's 1 cross section S is variable. It is equipped with a
push device to maintain a push force F in direction orthogonal to
atomic planes 2 of this monocrystal 1, located between distances
$R_0$ and $R_m$}
\end{figure}

\newpage

\begin{figure}[h]
\begin{center}
\includegraphics[width=4in]{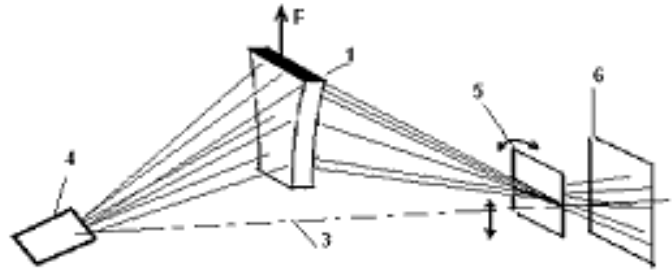}
\end{center}
\caption{Rotating X-ray scanning microscope consists of a source 4
of X-ray radiation, single monocrystal and the specimen 5. The
specimen 5 is after each revolution  shifted in direction of the
arrow, thus allowing its scanning from one end to the other.
Intensities of radiation measured by radiation detector 6 are
processed by computer into the resulting image similarly as in case
of tomography. Simultaneously, this monocrystal 1 acts a
monochromator and a separate monochromator may thus be excluded.}
\end{figure}

\newpage

\begin{figure}[h]
\begin{center}
\includegraphics[width=4in]{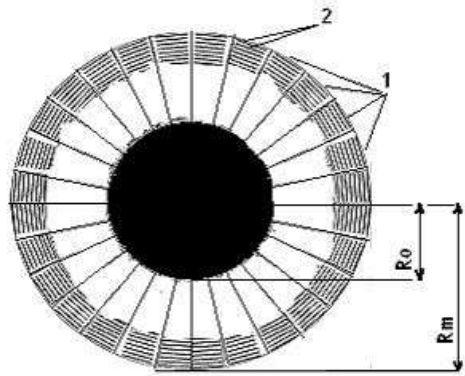}
\end{center}
\caption{Optical element consisting of 28 monocrystals }
\end{figure}

\newpage

\begin{figure}[h]
\begin{center}
\includegraphics[width=4in]{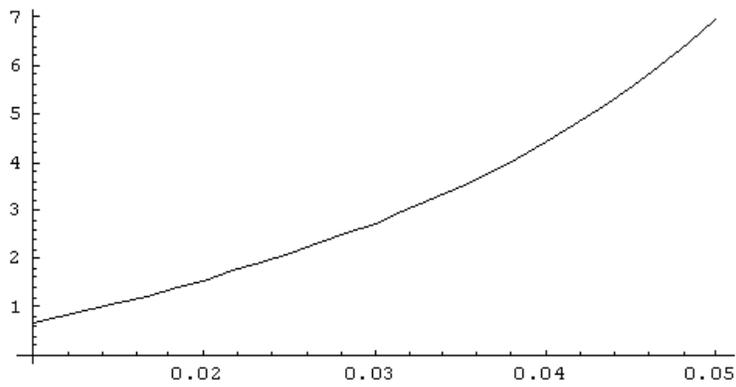}
\end{center}
\caption{dependencies of sections ratios $S/S_0$ of monocrystal 1 on
distance from optical axis 3 used for determination of pull force on
atomic planes 2 for the following values: $\lambda$ = 0.1 nm, $
d_{0}$ = 0.4 nm, $R_{0}$ = 0.01 m, s = 0.5 m and n = 1, in the
extent up to $R_{m}$ = 0,07 m.}
\end{figure}

\end{document}